\documentclass[10pt,a4paper,twoside]{article}
\usepackage{epsfig}
\usepackage{baltlat6}
\usepackage{array}
\usepackage{here}
\usepackage{natbib}
\usepackage{hyperref}

\makeatletter
\def\captionb#1#2{\refstepcounter{figure}\small\vspace{2mm}
\vbox{\baselineskip=9.5pt\hhuad{\bfseries Fig.\,#1.}
 \,#2} \baselineskip=11pt}

\def\sectionb#1#2{\refstepcounter{section}\vspace{5mm}\hbox{\kern-.9pt
{#1.\ \ }\vtop{\noindent #2}\nopagebreak}
\vspace{1mm} \noindent\baselineskip=11pt}

\def\subsectionb#1#2{\refstepcounter{subsection}\vspace{3mm}\hbox{\kern-.9pt
{\it #1.\ \ }\vtop{\noindent\it #2}\nopagebreak}
\vspace{.5mm} \noindent\baselineskip=11pt}

\def\subsubsectionb#1#2{\refstepcounter{subsubsection}\vspace{3mm}\hbox{\kern-.9pt
{#1.\ \ }\vtop{\noindent #2}\nopagebreak}
\vspace{.5mm} \noindent\baselineskip=11pt}
\makeatother

\bibpunct{(}{)}{;}{a}{}{,}
\begin{document}
\ \ \vspace{0.5mm} \setcounter{page}{139} \vspace{8mm}

\titlehead{Baltic Astronomy, vol.\,25, 139--151, 2016}

\titleb{CLASSIFYING ORBITS OF LOW AND HIGH ENERGY STARS IN AXISYMMETRIC
DISK GALAXIES}

\begin{authorl}
\authorb{Euaggelos E. Zotos}{}
\end{authorl}

\moveright-3.2mm
\vbox{
\begin{addressl}
\addressb{}{Department of Physics, School of Science, Aristotle University of Thessaloniki, \\
GR-541 24, Thessaloniki, Greece; \ e-mail: evzotos@physics.auth.gr}
\end{addressl}}

\submitb{Received: 2015 May 1; accepted: 2015 June 26}

\begin{summary} The ordered or chaotic character of orbits of stars
moving in the meridional $(R,z)$ plane of an analytic axisymmetric
time-independent disk galaxy model with an additional spherically
symmetric central nucleus is investigated.  Our aim is to determine
how the total energy influences the orbital structure of the galaxy.
For this purpose we monitor how the percentage of chaotic orbits as
well as the rates of orbits composing the main regular families
evolve as a function of the value of the energy.  In order to
distinguish with certainty between chaotic and ordered motion we use
the SALI method in extensive sets of initial conditions of orbits.
Moreover, a spectral method is applied for identifying the various
regular families and also for recognizing the secondary resonances
that bifurcate from them.  Our numerical computations suggest that
for low energy levels the observed amount of chaos is high and the
orbital content is rather poor, while for high energy levels,
corresponding to global motion, regular motion dominates and many
secondary higher resonances emerge.  We also compared our results
with previous related work.  \end{summary}

\begin{keywords}
galaxies: kinematics and dynamics -- galaxies: structure, chaos
\end{keywords}

\resthead{Orbits of low and high energy stars in axisymmetric disk
galaxies} {Euaggelos E. Zotos}

\sectionb{1}{INTRODUCTION}
\label{intro}

A first step towards understanding the overall orbital content of a
galaxy model is to know whether the orbits of stars are ordered or
chaotic. Another issue of significant importance is the distribution
of ordered orbits into different families.  For this particular
purpose, more than thirty years ago, \citet{BS82,BS84} introduced a
method based on spectral dynamics.  This technique was improved and
extended later on by \citet{L93} and \citet{CA98}.  The main
advantage of this method is the calculation of the ratio of the
integers which corresponds to the ratio of the main frequencies of
the orbit, where main frequency is the frequency of the greatest
amplitude in each coordinate.  The computation of the frequencies is
obtained by the frequency Fourier transform, a routine initially
developed by \citet{L88} and later substantially improved by
\citet{SN96}.  In this work, we shall deploy this method in order to
identify the resonant classification of regular orbits and
categorize them into different families.

Recently, in a series of papers \citep{CarZ13,ZCar13,Z14a,Z14b} we
classified orbits in several types of galaxy models containing dark
matter haloes.  These papers act as a guide regarding the structure
and the numerical methods of the present work.  Furthermore, in
other recent papers we performed orbit classification in new galaxy
models \citep{ZCar14}, or we investigated specific features of
well-known galactic systems, such as the influence of the angular
momentum \citep{Z14c} and the mass transportation from the disk to
the nucleus \citep{Z14d}.

In \citet{ZC13} (hereafter Paper I) we introduced a composite
analytic, axially symmetric galactic gravitational model that
embraces the general features of a disk galaxy with a dense massive
spherical nucleus.  Then we distinguished between regular and
chaotic motion of stars moving in the meridional $(R,z)$ plane.  We
also performed an orbit classification by separating regular orbits
into different families, thus revealing how
some important quantities of the system affect the orbital
structure.  In the present paper we shall use the same galactic
model in an attempt to determine the properties of low and high
energy stars.

The paper is organized as follows:  the properties of the galactic
gravitational model are presented in Section \ref{galmod}, while all
of the computational methods used for determining the character of
orbits are described in Section \ref{cometh}.  In the following
section we examine how the total energy influences the orbital
structure of the galaxy by measuring the corresponding percentages
of all types of orbits.  In Section \ref{over} we conduct an
overview analysis using a continuous spectrum of energy values.  The
article ends with Section \ref{disc}, where the conclusions of our
work are presented.

\sectionb{2}{PROPERTIES OF THE GALACTIC MODEL}
\label{galmod}

The aim of this numerical investigation is to classify the orbits of low
and high energy stars moving in the meridional $(R,z)$ plane of an
axisymmetric disk galaxy with an additional spherically symmetric
nucleus.  In our calculations we use cylindrical coordinates $(R, \phi,
z)$, where $z$ is the axis of symmetry of the galaxy.

Let us briefly recall the galactic gravitational model which was
introduced in Paper I. The total analytical gravitational potential
$\Phi(R,z)$ is time-independent and consists of two main components:
the potential of the central nucleus, $\Phi_{\rm n}$, and that of
the flat disk, $\Phi_{\rm d}$.  In order to describe the properties
of the nucleus we apply a spherically symmetric Plummer potential
(see, e.g., \citealp{BT08})
\begin{equation}
\Phi_{\rm n}(R,z) = - \frac{G M_{\rm n}}{\sqrt{R^2 + z^2 + c_{\rm n}^2}},
\label{Vn}
\end{equation}
where $G$ is the gravitational constant, while $M_{\rm n}$ and
$c_{\rm n}$ are the mass and the scale length of the nucleus,
respectively.  It should be pointed out that similar types of
potentials have been successfully used in previous works to model
the properties of central mass concentrations in galaxies (see,
e.g., \citealp{HN90,HPN93,Z12,ZC13,Z14a}).  Moreover, we would like
to emphasize that the central nucleus represents a bulge rather than
a black hole or any other compact stellar object and therefore we do
not take into account any relativistic effects.

To represent the flat galactic disk we use the potential of the
\citet{MN75} form:

\begin{equation}
 \Phi_{\rm d}(R,z) = - \frac{G M_{\rm d}}{\sqrt{R^2 + \left(\alpha +
\sqrt{h^2 + z^2}\right)^2}}\,.
\end{equation}
Here $M_{\rm d}$ is the mass of the disk, $\alpha$ is the horizontal
scale length of the disk and $h$ corresponds to the vertical scale
length of the disk.

In our work we use a system of galactic units where $G = 1$, the
unit of length is 1~kpc and the unit of velocity is 10 km\,s$^{-1}$.
Thus, the unit of time is $0.9778 \times 10^8$ yr, the unit of mass
is $2.325 \times 10^7\,M_{\odot}$, the unit of angular momentum (per
unit mass) is 10 km\,kpc\,s$^{-1}$, and the unit of energy (per unit
mass) is 100 km$^2$s$^{-2}$.  In these units, the values of the
involved parameters are:  $M_{\rm n} = 500$ (corresponding to 1.162
$\times 10^{10}\,M_{\odot}$), $c_{\rm n} = 0.25$, $M_{\rm d} = 7000$
(corresponding to 1.627 $\times 10^{11}\,M_{\odot}$), $\alpha = 3$,
and $h = 0.175$. The values of the nucleus and the disk were chosen
with a Milky Way-type galaxy in mind (e.g., \citealp{AS91}).
Furthermore, the above-mentioned values of the parameters are kept
constant throughout the numerical calculations and define the
Standard Model (S.M.) which secures everywhere a positive mass
density and is free of singularities.

 Due to the fact that the disk galaxy is axially symmetric, the $z$
 component of the total angular momentum is conserved. Exploiting this
property we can describe orbits by means of the effective potential
\begin{equation}
\Phi_{\rm eff}(R,z) = \Phi(R,z) + \frac{L_z^2}{2R^2}.
\label{veff}
\end{equation}
  The value of the angular momentum of all orbits is $L_{\rm z} = 10$
and remains constant throughout.

  In this case, the set of equations of motion in the meridional plane
is
\begin{equation}
\ddot{R} = - \frac{\partial \Phi_{\rm eff}}{\partial R}, \ \ \ \ddot{z} = - \frac{\partial \Phi_{\rm eff}}{\partial z},
\label{eqmot}
\end{equation}
while the variational equations needed for the calculation of chaos
indicators (in this case the SALI, see \S3) are
\begin{eqnarray}
 \dot{(\delta R)} &=& \delta \dot{R}, \ \ \ \dot{(\delta z)} = \delta
\dot{z}, \nonumber \\
(\dot{\delta \dot{R}}) &=&
- \frac{\partial^2 \Phi_{\rm eff}}{\partial R^2} \delta R
- \frac{\partial^2 \Phi_{\rm eff}}{\partial R \partial z}\delta z,
\nonumber \\
(\dot{\delta \dot{z}}) &=&
- \frac{\partial^2 \Phi_{\rm eff}}{\partial z \partial R} \delta R
- \frac{\partial^2 \Phi_{\rm eff}}{\partial z^2}\delta z.
\label{vareq}
\end{eqnarray}

The corresponding Hamiltonian to the effective potential
(\ref{veff}) reads
\begin{equation}
H(R,z,\dot{R},\dot{z}) = \frac{1}{2} \left(\dot{R}^2 + \dot{z}^2 \right) + \Phi_{\rm eff}(R,z) = E,
\label{ham}
\end{equation}
where $\dot{R}$ and $\dot{z}$ are momenta per unit mass, conjugate
to coordinates $R$ and $z$, respectively, while $E$ is the numerical
value of the Hamiltonian, which is conserved.  Consequently, a
trajectory is restricted to the area in the $(R,z)$ plane,
satisfying $E \geq \Phi_{\rm eff}(R,z)$, while all the other regions
are forbidden to the star motion.

\sectionb{3}{COMPUTATIONAL METHODS}
\label{cometh}

For the investigation of the orbital dynamics (regularity or
chaoticity) of the galaxy model, we need to establish some sets of
initial conditions of orbits.  For this purpose we define, for
several values of energy $E$, dense uniform grids of $1024 \times
1024$ initial conditions in the phase space $(R,\dot{R})$  regularly
distributed in the area allowed by the value of the Hamiltonian $E$.
The initial conditions of the orbits whose properties will be
examined are defined as follows:  we consider orbits with the
initial conditions $(R_0, \dot{R_0})$ with $z_0 = 0$, while the
value of $\dot{z_0}$ is always obtained from the Hamiltonian
(\ref{ham}). For each initial condition a double precision
Bulirsch-Stoer \verb!FORTRAN 77! algorithm (e.g., \citealp{PTVF92})
is used for the numerical integration of the equations of motion
(\ref{eqmot}) as well as the variational equations (\ref{vareq}).
The time step of the numerical integration is $10^{-2}$ and our
tests suggest that the results do not change by halving the time
step.  In all calculations the Hamiltonian (Eq.~\ref{ham}) was
conserved better than one part in $10^{-11}$, although for the
majority of the initial conditions it was better than one part in
$10^{-12}$.

The phase space is divided into the ordered and chaotic space.  Over
the years, several chaos indicators have been developed in order to
determine the character of orbits.  In our case, we chose to use the
Smaller ALingment Index (SALI) method.  The SALI \citep{S01} has
proved a very fast, reliable and effective tool, which is defined as
\begin{equation}
\rm SALI(t) \equiv min(d_-, d_+),
\label{sali}
\end{equation}
where $d_- \equiv \| {\bf{w_1}}(t) - {\bf{w_2}}(t) \|$ and $d_+
\equiv \| {\bf{w_1}}(t) + {\bf{w_2}}(t) \|$ are the alignments
indices, while ${\bf{w_1}}(t)$ and ${\bf{w_2}}(t)$ are two deviation
vectors which initially point in two random directions.  For
distinguishing between ordered and chaotic motion, all we have to do
is to compute the SALI along time interval $t_{\rm max}$ of
numerical integration.  In particular, we track simultaneously the
time-evolution of the main orbit itself as well as the two deviation
vectors ${\bf{w_1}}(t)$ and ${\bf{w_2}}(t)$ in order to compute the
SALI.  The time-evolution of SALI strongly depends on the nature of
the computed orbit: in the case of a regular orbit, the SALI
exhibits small fluctuations around non-zero values, but when the
orbit is chaotic, after a small transient period the SALI tends
exponentially to zero, approaching the limit of the accuracy of the
computer $(10^{-16})$.  Therefore, the particular time-evolution of
the SALI allows us to distinguish fast and safely between regular
and chaotic motion.  Nevertheless, we have to define a specific
numerical threshold value for determining the transition from order
to chaos. After conducting extensive numerical experiments,
integrating many sets of orbits, we conclude that a safe threshold
value for the SALI is $10^{-8}$.  In order to decide whether the
orbit is regular or chaotic, one may follow the usual method
according to which we check, after a certain and predefined time
interval of numerical integration, if the value of SALI becomes less
than the established threshold value. Therefore, if SALI $\leq
10^{-8}$, the orbit is chaotic, while when SALI $ > 10^{-8}$ the
orbit is regular, thus making the distinction between regular and
chaotic motion clear and beyond any doubt.  For the computation of
SALI we used the \verb!LP-VI! code \citep{CMD14}, a fully
operational routine which efficiently computes a suite of many chaos
indicators for dynamical systems in any number of dimensions.

The total integration time is an element of paramount importance. In
our case, each initial condition is integrated for $10^4$ time units
($10^{12}$ yr) which correspond to a time span of the order of
hundreds of orbital periods.  We have chosen to apply such a large
time interval having in mind the occasion of ``sticky"
orbits\footnote{~Usually a ``sticky" orbit behaves regularly for
long time periods before it finally drifts away from the boundaries
of ordered regions and starts to wander in the chaotic domain, thus
fully revealing its true chaotic nature.}.  Here it should be
emphasized that a sticky orbit could be easily misclassified as
ordered by any chaos indicator if the integration period is too low,
so that the orbit does not have enough time to unveil its chaotic
character.  Setting the integration interval to $10^{4}$ we manage
to correctly classify orbits with sticky periods of at least 100
Hubble times.  Orbits with higher sticky periods do not have any
physical meaning and are completely out of the scope of the present
work.

Before closing this section, we should better clarify an issue
regarding the nomenclature of the orbits.  It is well known that in
axisymmetric potentials all orbits are in fact three-dimensional
loop orbits that always circulate in the same direction around the
$z$-axis of symmetry. In the meridional plane, however, the meaning
of the rotation is lost and therefore the path that an orbit follows
onto the $(R,z)$ plane can take any shape.  In this work, as in all
previous related papers, we characterize an orbit by taking into
account its motion in the meridional plane.  For instance, if the
orbit is a rosette in the 3D system, it will be a linear orbit in
the $(R,z)$ plane, a closed tube orbit will be a meridional 2:1
banana-type orbit, etc.  Furthermore, it should be emphasized that
the term ``box" orbit is used for the orbit that conserves
circulation but this refers {\bf only} to the circulation provided
by the meridional plane itself.  These orbits were originally
characterized as ``boxes" (e.g., \citealp{O62}) due to their boxlike
shape, even though their shapes in three-dimensions are more similar
to doughnuts (see the review of \citealp{M99}).  In order to
maintain the continuity with all previous related works we decided
to keep this initial formalism.

\sectionb{4}{ORBIT CLASSIFICATION}
\label{orbclas}

In this section we will investigate the orbital content of the
galaxy for several energy levels.  We use the sets of initial
conditions of orbits analyzed in Section \ref{cometh} in order to
construct the respective grids in the phase plane.  It should be
emphasized that the energy level controls the actual size of the
grid and particularly the $R_{\rm max}$ which is the maximum
possible value of the $R$ coordinate. We choose the values of the
energy which yield 3 kpc $< R_{\rm max} <$ 18 kpc.  Once the value
of the energy is chosen and the corresponding grid is determined, we
integrate the initial conditions of orbits, calculate the SALI and
then classify ordered orbits into regular families.

Our numerical calculations suggest that in our axially symmetric
galaxy model there are eight basic types of orbits:  (i) chaotic
orbits; (ii) box orbits; (iii) 1:1 linear orbits; (iv) 2:1
banana-type orbits; (v) 2:3 fish-type orbits; (vi) 4:3 resonant
orbits; (vii) 6:5 resonant orbits and (viii) orbits with other
secondary resonances (i.e., all resonant orbits not included in the
former categories).  It turns out that for each resonant family
included in the ``other" category the corresponding percentage is
less than 1\% in all examined cases, and therefore their
contribution to the overall orbital structure of the galaxy is
practically insignificant.  The $n :  m$ notation\footnote{~An $n:m$
resonant orbit would be represented by $m$ distinct islands of
invariant curves in the $(R,\dot{R})$ phase plane and $n$ distinct
islands of invariant curves in the $(z,\dot{z})$ surface of
section.} we use for the regular orbits is according to \citet{CA98}
and \citet{ZC13}, where the ratio of these integers corresponds to
the ratio of the main frequencies of the orbit, where main frequency
is the frequency of the greatest amplitude in each coordinate.  Main
amplitudes, when having a rational ratio, define the resonances of
an orbit.  The shapes of all of the basic types of orbits are shown
in Fig.\,3 of Paper I.

\begin{figure*}[!tH]
\centerline{ \resizebox{\hsize}{!}{\includegraphics{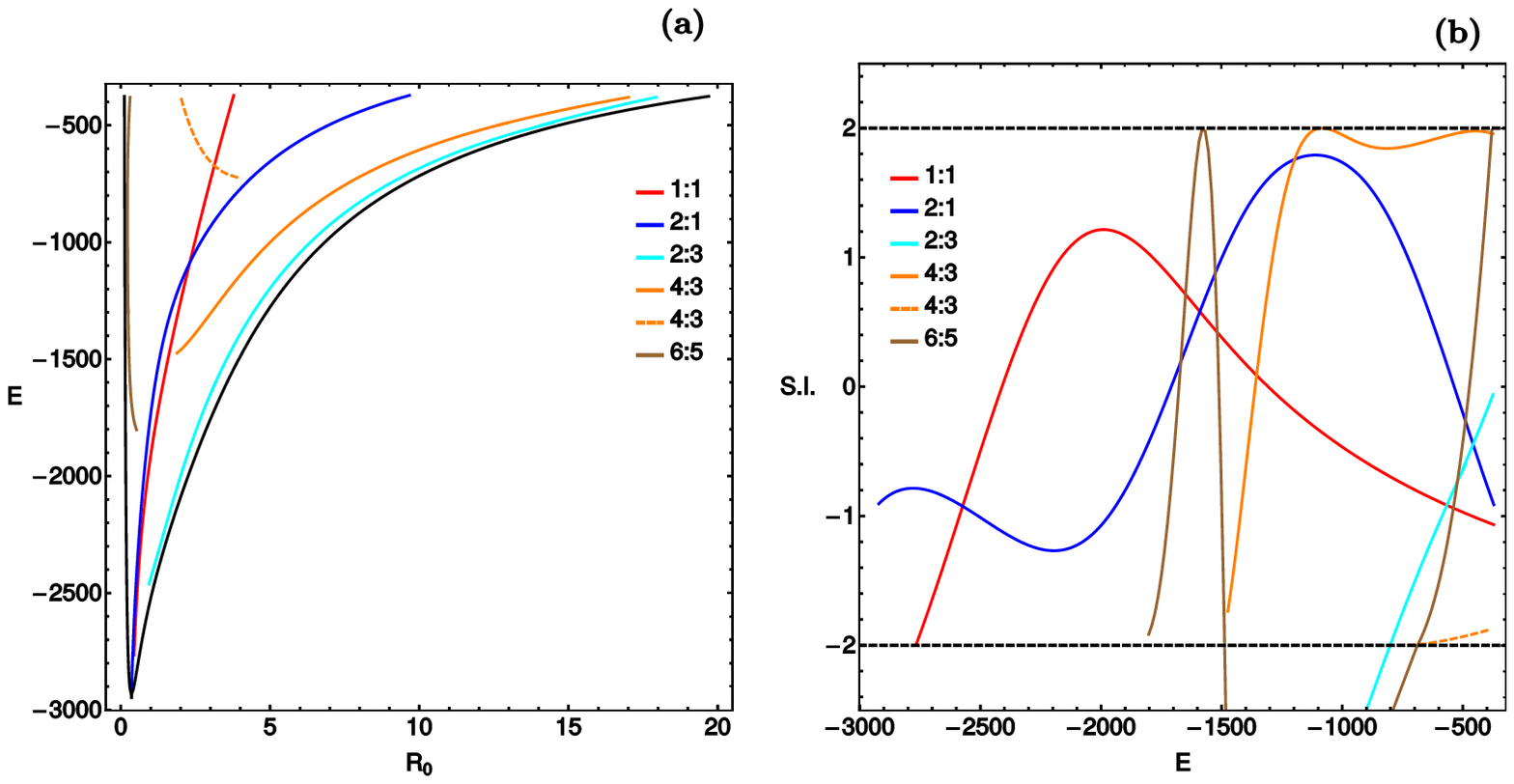}}}
\captionb{1}{Panel (a):  the $(R_0, E)$ characteristic curves of all
the orbital families. Panel (b):  the evolution of the stability
index (S.I.) of all the families of periodic orbits shown in panel
(a) as a function of the orbital energy $E$.  The black horizontal
dashed lines at --2 and +2 delimit the range of S.I. for which the
periodic orbits are stable.} \label{fpos}
\end{figure*}

Panel (a) of Fig.\,1 presents the so-called ``characteristic"
diagram which shows the evolution of the initial condition $R_0$ of
the parent periodic orbits of each resonant family as a function of
the total orbital energy $E$.  Here we would like to point out that
for orbits launched perpendicular to the $R$-axis only the initial
condition $R_0$ is needed to locate them on this informative
diagram. For orbits not starting perpendicularly to the $R$-axis
(i.e., the 1:1 family), we need pairs of initial position-velocity
$(R_0,\dot{R_0})$ and, therefore, the characteristic diagram is now
three-dimensional.  In the two-dimensional characteristic diagrams
of panel (a) we observe that the curve of the 1:1 resonant family
crosses two other characteristic curves.  Here it should be
clarified that the full characteristic curve of the 1:1 resonant
family is three-dimensional since $\dot{R_0} \neq 0$, however we
decided to combine all families together in a two-dimensional plot
containing only the evolution of the $R$ coordinate (of course, in
the 3D $(R, \dot{R}, E)$ space the corresponding characteristic
curves do not cross one another).  The outermost black solid line
denotes the permissible area of motion inside the potential well
$\Phi_{\rm eff}(R, z = 0)$.  In Fig.\,1b, we present the so-called
``stability" diagram which indicates the stability of all of the
resonant families as the numerical value of the energy varies.  This
plot gives us the ability to examine the evolution of the stability
index (S.I.) \citep{MH92,Z13} of the resonant periodic orbits as a
function of energy and also to monitor all the transitions from
instability to stability and vice versa.  We shall recall that a
periodic orbit is stable if S.I. is between --2 and +2.  Our
calculations reveal that the 2:1 resonant family is stable
throughout the energy range $(E \in [-2946, -310])$. The 1:1
resonant family  starts at $E = -2764.88$ and remains stable
throughout. Moreover, the 2:3 resonant family begins at $E =
-2462.07$ and is unstable in the interval $-2462.07 \leq E \leq
-804.71$.  Similarly, the starting point of the 4:3 resonant family
is at $E = -1474.07$. Looking carefully at panels (a) and (b) we see
that there is a second smaller and stable branch of the 4:3 resonant
family (shown as a dashed line) for $-722.12 \leq E \leq -371$.
Finally, the 6:5 resonant family initiates at $E = -1802.01$ and is
unstable only in the interval $-1359.24 \leq E \leq -686.91$.


\begin{figure}[!tH]
\centerline{ \resizebox{0.81\hsize}{!}{\includegraphics{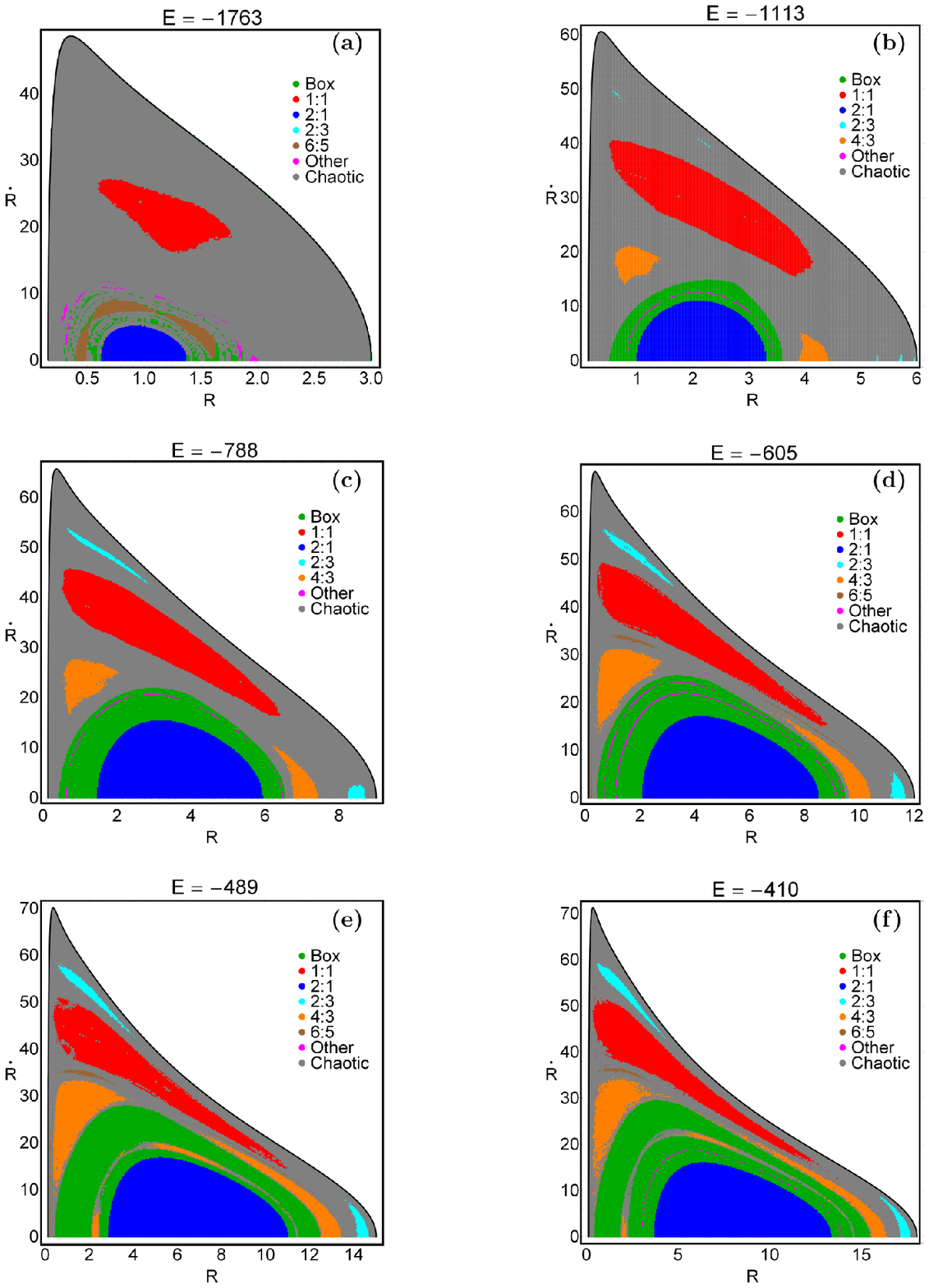}}}
  \captionb{2}{Orbital structure of the phase plane $(R,\dot{R})$ of our
disk galaxy model for different values of the total orbital energy
$E$.} \label{grd}
\end{figure}

Fig.\,2 with panels from (a) to (f) presents six grids of initial
conditions $(R_0,\dot{R_0})$ of orbits that we have classified for
different values of the orbital energy.  These color-coded grids are
equivalent to the classical Poincar\'{e} Surface of Section (PSS)
and help us to determine what types (families) of orbits occupy
specific domains in the phase  plane $(R,\dot{R})$.  The outermost
thick black curve is the limiting curve in the phase space which is
defined as
\begin{equation}
\frac{1}{2} \dot{R}^2 + \Phi_{\rm eff}(R,z = 0) = E.
\label{zvc}
\end{equation}

Panel (a) of Fig.\,2 shows the grid for a very low energy level.
This level corresponds to local motion around the dense and massive
nucleus. Therefore, we observe that the vast portion of the plane is
covered by initial conditions of chaotic orbits.  There are two main
stability domains corresponding mainly to 2:1 and 1:1 resonant
orbits.  Around the 2:1 stability island we identify delocalized
initial conditions of box orbits and a chain of 6:5 stability
islands.  In addition, above the 6:5 stability islands we observe
the presence of scattered initial conditions corresponding to higher
order resonant orbits (mainly 6:7 and 8:7 resonant orbits).  In
Fig.\,2b the energy level is $E = -1113$. Here, several differences
with respect to the previous grid are observable:  (i) the amount of
chaos has decreased, while the stability islands of 2:1 and 1:1
resonant orbits occupy a greater area in the phase plane; (ii) the
initial conditions of box orbits form an open ring above the 2:1
stability island; (iii) the stability islands of the 4:3 resonance
have emerged; (iv) some initial conditions corresponding to 2:3
resonant orbits are identified inside the unified chaotic sea. The
stability diagram shown in Fig.\,1b indicates that the 2:3 resonance
is unstable for $E = -1113$, therefore it is impossible to form
well-defined stability islands like the other resonances.  Inside
the box region there is a stripe of initial conditions that
correspond to the 7:5 resonance.  As the energy increases, the
extent of the chaotic domain decreases.  This is clear in Fig.\,2c
where $E = -788$.  In this case, the 2:3 resonance is stable and
therefore we can see the two\footnote{~It must be emphasized that
the color-coded grids shown in Fig.\,\ref{grd}\,(a--f) present only
the $\dot{R} > 0$ part of the  phase plane $(R,\dot{R})$; the
$\dot{R} < 0$ part is symmetrical. Therefore, in many resonant cases
not all of the corresponding stability islands are shown (e.g., for
the 2:3 and 4:3 resonances only two (one and a half to be more
precise) of the total three islands are present).} corresponding
stability islands.  We should also mention that the 6:5 resonance is
absent in the grids of Figs.~2b and 2c.  This, however, is
anticipated because the 6:5 resonance is unstable for $-1359.24 \leq
E \leq -686.91$ (see Fig.\,1b) and therefore the corresponding
periodic point is deeply buried in the chaotic sea.  Fig.\,2d is
quite similar to Fig.\,2c, with only minor differences.  The 6:5
resonance emerges again, while in the box area there are two small
filaments of initial conditions corresponding to secondary
resonances (5:4 and 7:5). When $E = -489$, we observe in Fig.~\,2e
that the chaotic sea is reduced to a chaotic layer situated at the
outer parts of the phase plane.  We also see the presence of the
second chain of the 4:3 resonance inside the box region, while
secondary and higher resonant orbits seem to be absent in this case.
It should be pointed out that our result for $E = -489$ is quite
similar to that presented in Fig.\,18a of Paper I (the only
differences are due to improvements in the classification code).
Finally, in Fig.\,2f we present the orbital structure of the phase
plane for $E = -410$.  It is seen that the structure is very similar
to that Fig.\,2e discussed above. Inside the box region one can
identify tiny chains of initial conditions corresponding to either
5:4 or 9:7 resonance. For relatively high energy levels, or, in
other words, for global motion of stars, we see that the orbital
content remains almost the same regardless of the orbital energy
shift.


\begin{figure*}[!tH]
\centerline{ \resizebox{0.60\hsize}{!}{\includegraphics{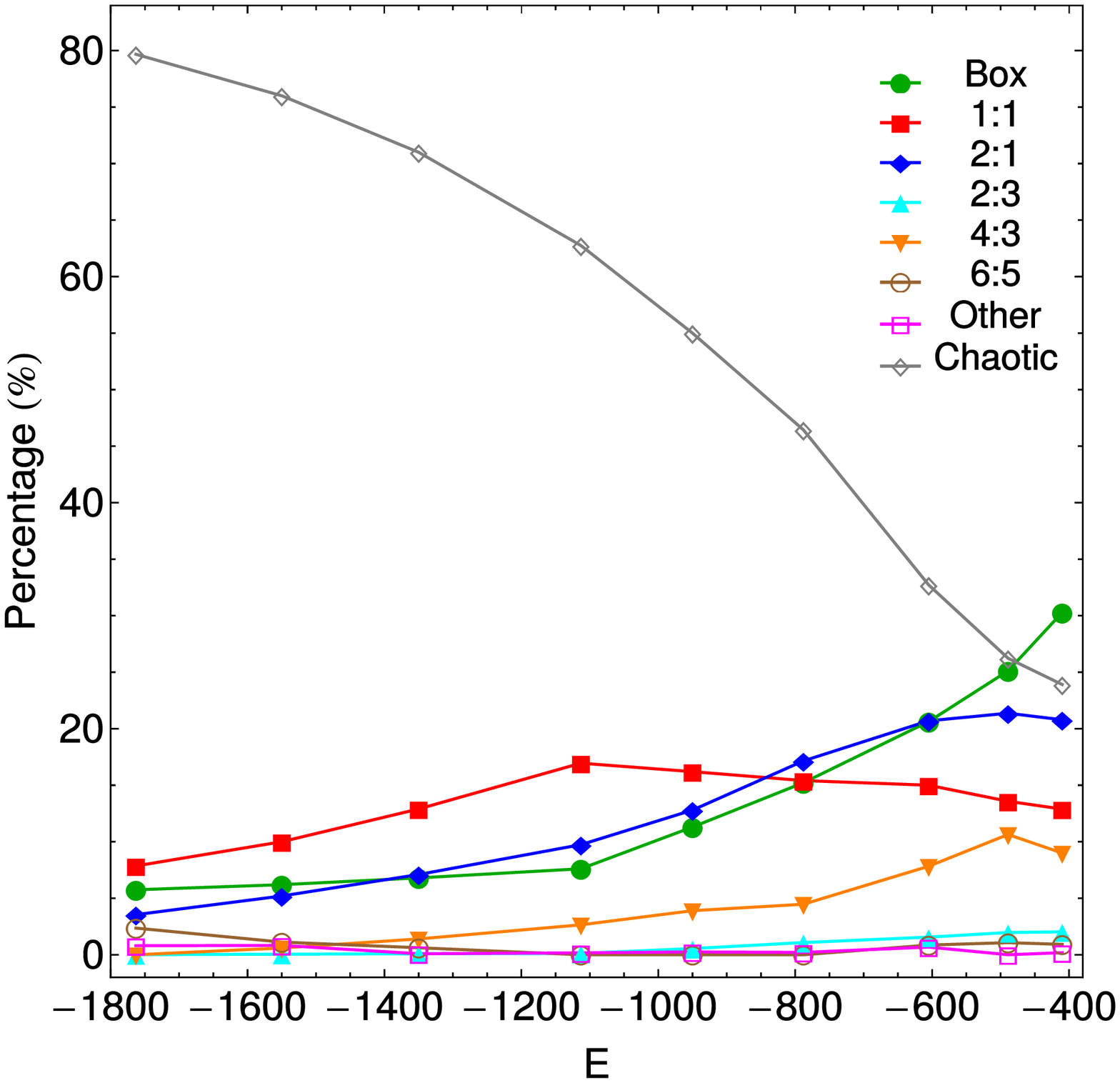}}}
\captionb{3}{Evolution of the percentages of different types of
 orbits in the phase plane $(R,\dot{R})$ of our galaxy model as a
function of the total orbital energy $E$.} \label{percs}
\end{figure*}

The evolution of the percentages of all types of orbits as a
function of the total orbital energy $E$ is shown in
Fig.\,\ref{percs}. We observe that for low energy levels,
corresponding to local motion around the central nucleus, chaotic
motion dominates which covers about 80\% of the entire phase plane.
As we proceed to higher energy levels however, the percentage of
chaos decreases constantly and, at the highest energy level studied,
chaos occupies only about 10\% of the phase space.  On the other
hand, the rate of many regular families grows with increasing
energy. In particular, the percentage of box orbits exhibits an
almost exponential increase and for $E = -410$ it is the most
populated family, implying that in global motion the majority of the
stars perform box orbits.  The percentage of 2:1 resonant orbits
also increases, however for $E > -600$ it seems to saturate around
20\%. Furthermore, the percentages of 1:1 and 4:3 resonant orbits
initially grow, but they drop for $E > -1100$ and $E > -500$,
respectively. Moreover, we could say that, in general terms, all the
other resonant families (2:3, 6:5 and other) possess throughout very
low percentages (always less than 3\%), thus varying the value of
the energy only shuffles the orbital content among them. Therefore,
taking into account all the above-mentioned numerical analysis, we
may conclude that in the phase space $(R,\dot{R})$  the types of
orbits that are mostly influenced by the value of the orbital energy
are box, 1:1, 2:1, 4:3 and chaotic orbits.  In general terms, the
numerical outcomes presented in Fig.\,\ref{percs} coincide with
those reported in Fig.\,18b of Paper I.

\sectionb{5}{AN OVERVIEW ANALYSIS}
\label{over}


\begin{figure*}[!tH]
\centerline{
\resizebox{0.60\hsize}{!}{\includegraphics{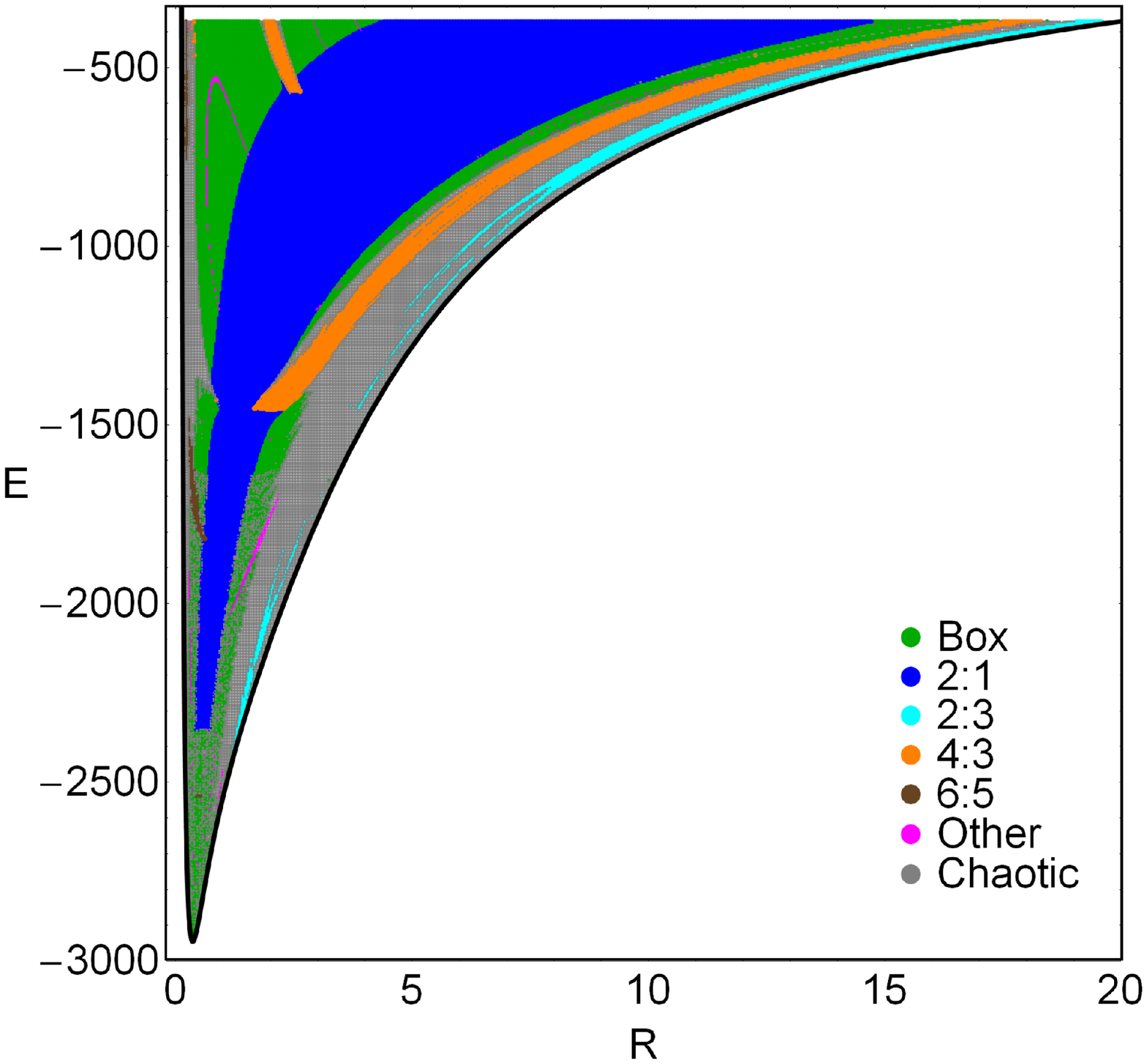}}}
 \captionb{4}{Orbital structure of the $(R,E)$ plane showing a detailed
 analysis of the evolution of orbits starting perpendicularly to the
 $R$-axis when the value of energy varies in the interval $E \in
[-2946, -371]$.}
\label{clas}
\end{figure*}

\begin{figure*}[!tH]
\centerline{
\resizebox{0.60\hsize}{!}{\includegraphics{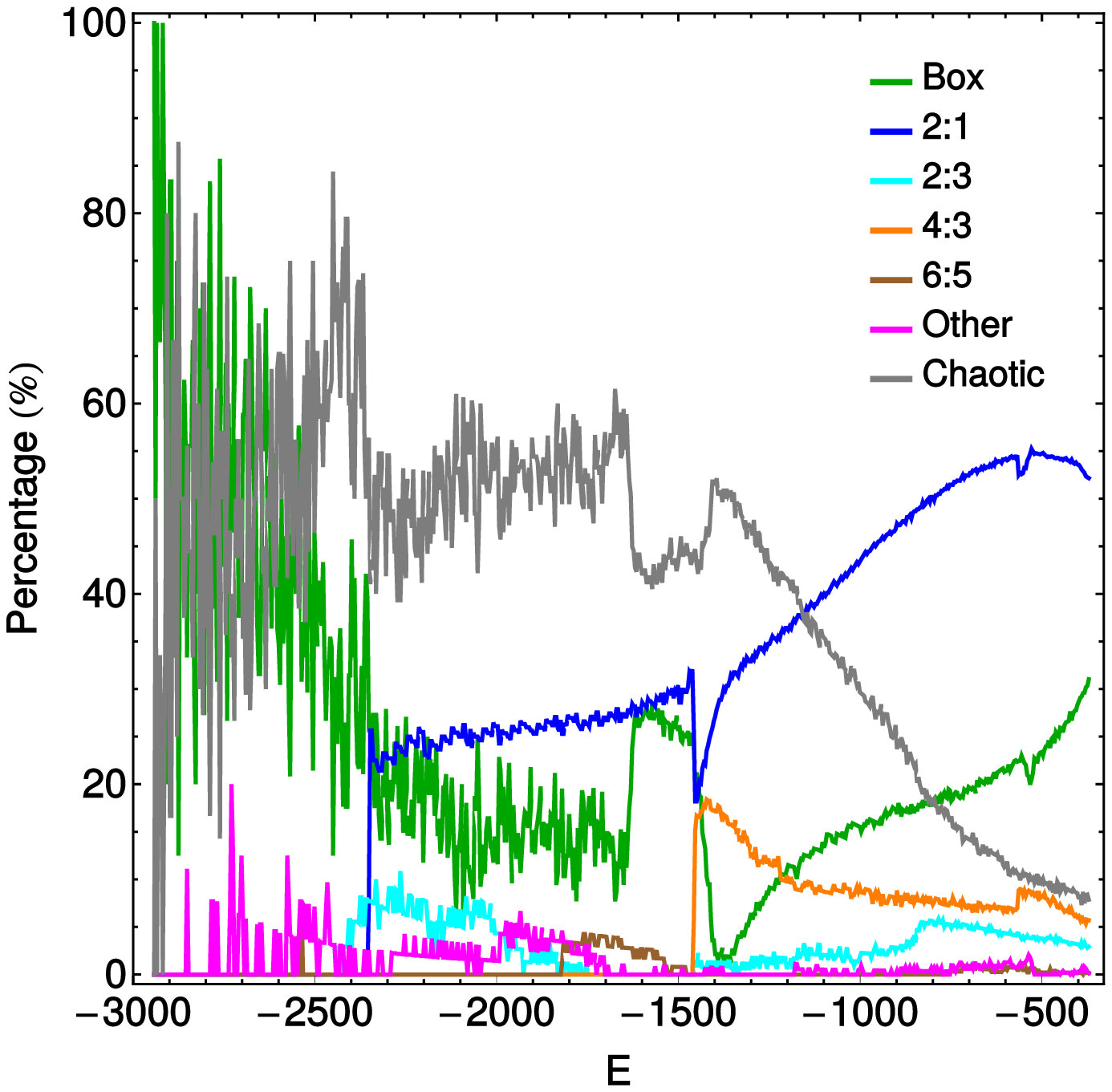}}}
 \captionb{5}{Evolution of the percentages of different types of
orbits in the $(R,E)$ plane of our galaxy model as a function of the
total orbital energy $E$.} \label{percs2}
\end{figure*}

The color-coded grids in the phase space $(R,\dot{R})$  presented in
Figs.\,2a--2f can provide information on the phase space mixing, but
only for a fixed value of the energy integral.  Back in the late 60s
H\'{e}non conceived a new type of plane using the section $z =
\dot{R} = 0$, $\dot{z} > 0$.  In this case, all the stars are
launched from the horizontal $R$ axis, parallel to the vertical $z$
axis and in the positive direction.  This plane provides also
information regarding areas of regularity and chaotic domains, but,
in contrast to the previously described type of grids, only orbits
with pericenters on the horizontal axis are included.  Therefore,
the value of the Hamiltonian $E$ can be used as an ordinate giving
us the ability to investigate a continuous spectrum of energy values
rather than a few discrete energy levels.  In Fig.\,\ref{clas} we
present the orbital structure of the $(R,E)$ plane when $E \in
[-2946, -371]$.  We clearly distinguish several stability regions of
resonant islands and chaotic layers.  Once more, in order to explore
with sufficient accuracy the evolution of the types of orbits, we
defined dense uniform grids of $1024 \times 1024$ initial conditions
in the $(R,E)$ space.  The plot shown in Fig.\,\ref{clas} reminds us
a lot of the characteristic diagram presented earlier in
Fig.\,\ref{fpos}a.  At high energy levels, we observe that inside
the box region, several families of other higher resonances are
present producing thin filaments of initial conditions.

In addition, in Fig.\,\ref{percs2} we provide a diagram showing the
evolution of the percentages of all types of orbits as a function of
the energy.  This diagram is quite similar to that presented in
Fig.\,\ref{percs} but in this case each curve is composed of a
continuous spectrum of the values of energy.  Here it should be
pointed out that the $(R,E)$ plane contains only such orbits which
start perpendicularly to the $R$ axis.  Therefore, the 1:1 resonant
family for which the initial conditions are pairs of
position-velocity is obviously not included.  It is evident that,
for very low values of energy $(E < -2500)$, the $(R,E)$ plane is
dominated mainly by a mixture of box and chaotic orbits.  For
intermediate energy levels $(- 2500 < E < -1500)$, we see that the
percentages of all types of orbits exhibit minor fluctuations, thus
remaining almost constant. For higher energy levels $(E > -1500)$,
however, we observe that the increase on the energy influences, more
or less, the populations of all types of orbits. In particular, the
percentage of chaotic orbits decreases from about 50\% to about
10\%, while at the same time the rates of box and 2:1 resonant
orbits grow almost linearly and at the highest energy level studied
$(E = -371)$ they occupy about 30\% and 55\% of the $(R,E)$ plane,
respectively. Furthermore, the percentages of the 2:3 and 4:3
resonant families seem to saturate around 5\%, while the rates of
the 6:5 family and other types of orbits remain very low (around
1\%).

\sectionb{6}{DISCUSSION AND CONCLUSIONS}
\label{disc}

In this work we adopted the analytic axisymmetric galactic
gravitational potential introduced in Paper I and attempted to
explore how the value of the total orbital energy influences the
level of chaos and also the distribution of the main regular
families.  Our numerical investigation took place in the meridional
$(R,z)$ plane, in order to reduce the three dimensional motion into
two dimensions. Moreover, the values of all the other dynamical
quantities entering the system were kept constant throughout so as
to monitor the evolution of the percentages of all types of orbits
as a function of the energy.  Our detailed and systematic numerical
investigation revealed that the populations of the
 regular families as well as the amount of chaos are indeed very
dependent on the value of the energy.

For a better understanding of the orbital properties of the galactic
system we examined the structure of the phase space $(R,\dot{R})$.
For distinguishing between order and chaos we defined dense grids of
$1024 \times 1024$ initial conditions of orbits uniformly
distributed within the area defined by the corresponding energy
level.  Then, for illustrating how the energy influences the orbital
structure, we presented in each case color-coded grids, thus
visualizing what types of orbits occupy specific domains of the
phase space.  Each initial condition was numerically integrated for
a time period of $10^4$ time units which corresponds to $10^{12}$
yr.  Our choice regarding the integration time allowed us to
eliminate sticky orbits with periods of at least 100 Hubble times.
Initially our numerical code used the SALI method to identify the
regular or chaotic nature of the orbits and then a frequency
analysis technique was applied in order to classify regular orbits
into different families.

The most important results of our numerical investigation can be
summarized as follows:
\begin{itemize}
\item The SALI has been proved, once more, a very reliable dynamical
indicator which allowed us to safely distinguish between regular and
chaotic initial conditions of orbits relatively fast and with great
accuracy.
\item In the phase space of our composite gravitational model, several
regular types of orbits were found to exist, while strong and extended
chaotic areas separating the stability domains were also present.  In
particular, a wide plethora of resonant orbits (i.e., 1:1, 2:1, 2:3,
4:3, 6:5 and other higher resonant orbits) have been identified thus
making the orbital content of the system more rich.
\item We revealed that in the phase space $(R,\dot{R})$  the box, 1:1,
2:1, 4:3 and chaotic orbits are those mainly influenced by the
change in the value of the energy.  Furthermore, it was observed
that most of the resonant families are stable, although they were
found in specific energy ranges in which some resonant families
(i.e., the 2:3 and 6:5) become unstable.
\item We examined the orbital content of several isolated energy levels.
It was found that for low energy levels that correspond to local
motion around the dense and massive spherical nucleus the motion of
stars is highly chaotic, but as the value of the energy increases,
leading to global motion, the amount of the observed chaos gradually
decreases and regular motions take over the phase space.
\item We proceeded one step further constructing the $(R,E)$ plane so as
to monitor the evolution of all types of orbits as a function of the
energy. It is found that at very low energies there is a strong
mixture of box and chaotic orbits which, however, dissolves as the
energy increases, thus giving room to many other types of regular
resonant orbits to grow.
\end{itemize}

Judging by the detailed outcomes we may say that our task has been
successfully completed.  We hope that the present analysis and the
corresponding numerical results will be useful in the field of orbit
classification in galaxy models.  Taking into account that our
results are encouraging, it is in our future plans to properly
modify our dynamical model in order to expand our investigation into
three dimensions and explore the entire six-dimensional phase space
thus revealing the influence of the value of the energy on the
orbital structure.\enlargethispage{-5mm}

\thanks{ We would like to express our warmest thanks to Dr.  Leonid P.
Ossipkov (University of Petersburg) for the careful reading of the
manuscript and for all suggestions and comments which allowed us to
improve both the quality and the clarity of our paper.}

\References \vspace*{1\baselineskip}

\begingroup
\renewcommand{\section}[2]{}

\endgroup

\end{document}